# Ligand-metal covalency effects in resonance enhanced x-ray Bragg diffraction

S. W. Lovesey[1, 2] and G. van der Laan[2]

[1]ISIS Facility, STFC, Didcot, Oxfordshire OX11 0QX, UK

[2]Diamond Light Source, Harwell Science and Innovation Campus, Didcot, Oxfordshire OX11 0DE, UK

**Abstract** Chlorine covalently bonded to an open shell metal is present in many materials with desirable or intriguing physical properties. Materials include highly luminescent nontoxic alternatives to lead halide perovskites for optoelectronic applications $K_2CuCl_3$ and $Rb_2CuCl_3$, enantiomorphic $CsCuCl_3$ that presents magneto-chiral dichroism at a low temperature, and cubic $K_2RuCl_6$ that possesses a singlet ground state generated by antiparallel spin and orbital angular momenta. Structural chirality of $CsCuCl_3$ has been confirmed by resonant x-ray Bragg diffraction. We explore likely benefits of the technique at the chlorine K-edge using a symmetry inspired method of calculation applied to chlorine multipoles. Already, a low energy feature in corresponding x-ray absorption spectra of many compounds has been related to the chlorine-metal bond. Bragg diffraction from chlorine in cubic $K_2RuCl_6$ is treated in detail. Diffraction patterns for rhombohedral compounds that present space-group forbidden Bragg spots are found to be relatively simple.

## I. INTRODUCTION

Covalent bonding can leave a bold imprint on x-ray absorption spectra. Notably, a pre-edge feature in spectra attributed to an admixture of ligand and metal orbitals has been observed. Such is the case in extensively investigated K-edge absorption spectra of Cl bound to an open shell metal [1, 2, 3, 4]. A significant pre-edge feature is due to a forbidden $1s \rightarrow 3d$ transition, which becomes partially allowed due to mixing of ligand p-orbitals with metal d-orbitals. This pre-edge feature is not present when chlorine is not ligated to an open shell transition metal [1]. The intensity of feature is dictated by the degree of covalency in the metal-Cl bond and is therefore sensitive to the electronic structure of the metal and its oxidation state. Beyond, Cl p-orbitals centred around $-4$ eV with respect to the Fermi level in the density of states and strongly hybridized with Ru d-orbitals in cubic $K_2RuCl_6$ may be responsible for charge transfer excitations observed by resonant inelastic x-ray scattering [5]. Several methods in routine use in inorganic chemistry provide an experimental estimate of the covalency of a metal-ligand bond. Including, (i) magnetic measurements of ground-state properties and (ii) excited-state and relaxation effects. The first group includes g-values, metal hyperfine, and ligand super-hyperfine coupling constants obtained from electron paramagnetic resonance spectroscopy [6]. The second group includes absorption spectroscopy (ligand-to-metal charge transfer and d-d transitions) and photoelectron core-level spectroscopy [7]. We investigate the use of resonant x-ray Bragg diffraction which has not been exploited so far, to the best of our knowledge. To this end, we calculated diffraction patterns for $K_2RuCl_6$ allowing for enhancement of Bragg spot intensities from absorption at the Cl K-edge. The compound is chosen for the principal study in part because of a challenging motif of Cl ions, and in part because of its unusual magnetic state that continues to receive attention [5]. Related copper and iron halides are mentioned in Section IV. In all cases, electronic properties of Cl ions are encapsulated in atomic multipoles.

Experimental data on the pre-edge feature of the Cl K-edge are recorded in Table 1 and Fig. (2a) of references [1] and [4], respectively, for example. Very informative absorption spectra are those of isostructural structures hosting [$ZnCl_4$] and [$CuCl_4$] units that exhibit major activities at an energy ≈ 2827 eV [2]. Additionally, the copper but not the zinc compound shows an intense low energy peak at ≈ 2820 eV. As $Zn^{2+}$ has a filled $d^{10}$ shell, the lowest unoccupied energy levels are the 4s and 4p orbitals. Thus, its edge features derive primarily from the electric-dipole (E1) 1s → 4p transition. Whereas, $Cu^{2+}$ possesses a $d^9$ configuration with a hole in the d shell [6]. The observed pre-edge transition was assigned by the authors as a Cl 1s → Cu 3d transition not possible in the Zn compound because of the filled d subshell [2]. This assignment appeals to an admixture between Cu 3d and Cl 3p orbitals with weight ρ, say. A direct Cl 1s → Cu 3d transition is accomplished with an electric-quadrupole (E2) event that has a relative weight $(1 - \rho^2)^{1/2}$. The Cl 3p-metal d hybridization leads to holes in the Cl p-orbital and E1-E1 events in x-ray scattering. This is similar as for the case of the oxygen 2p band, in which case the O 3d continuum state is too high in energy to give E2 transitions at low energy. Another remark worth making is that the low-energy pre-peak will correspond to the lowest unoccupied states. This suggests indeed bonding states, where Cl 3p electron density has been transferred to the metal d band (giving holes in the Cl 3p band accessible by E1-E1). Chlorine ions in $K_2RuCl_6$ occupy sites devoid of inversion symmetry that forbids a polar E1-E2 absorption event. Our calculated axial E1-E1 and polar E1-E2 scattering amplitudes are radically different.

Strengths of the x-ray scattering amplitudes depend on atomic radial integrals that are different for E1 and E2 events. Table I contains estimates of relevant integrals obtained from a tried and tested atomic code written by Cowan [8]. E1-E1 and E1-E2 x-ray scattering amplitudes are proportional to $(1s\|R\|3p)^2$ and $[q(1s\|R\|3p)(1s\|R^2\|3d)]$, respectively [9]. The photon wavelength $\lambda \approx (12.4/\Delta)$ Å with $\Delta \approx 2.82$ keV for the Cl K-edge, and the wave vector $q = (2\pi/\lambda)$. Table I includes for comparison radial integrals for the Ti K-edge at $\Delta \approx 4.97$ keV, and Ni K-edge at $\Delta \approx 8.34$ keV [10]. Using the quoted estimates of radial integrals and energies of the respective K-edges one finds the E1-E2 amplitude for Cl exceeds the corresponding amplitude for Ni by a factor ≈ 6.1.

## II. CRYSTAL PROPERTIES

The crystal structure of $K_2RuCl_6$ is depicted in Fig. 1. It uses centrosymmetric $Fm\overline{3}m$ (No. 225, crystal class $m\overline{3}m$) with Ru ($Ru^{4+}$, $4d^4$), K and Cl ions in sites 4(a), 8(c) and 24(e), respectively [11]. Cell length $a \approx 9.737$ Å. Parity-even (axial, $\langle T^K \rangle$) and parity-odd (polar, $\langle U^K \rangle$) electronic multipoles are permitted at Cl sites. Spherical multipoles $\langle T^K \rangle$ and $\langle U^K \rangle$ have integer rank K [9] (Cartesian and spherical components of a dipole **R** = (x, y, z) are related by x = ($R_{-1}$ − $R_{+1}$)/√2, y = i($R_{-1}$ + $R_{+1}$)/√2, z = $R_0$). Angular brackets ⟨ ... ⟩ denote the time-average, or expectation value, of the enclosed tensor operator, i.e., multipoles of interest in Bragg diffraction occupy the electronic ground state. They can be calculated using standard tools of atomic physics given a suitable wavefunction [9]. Alternatively, multipoles can be estimated from a simulation of electronic structure [12]. Both magnitudes and symmetries of the electronic multipoles depend on ligand-metal covalency.

Chlorine multipoles are prescribed by tetragonal 4mm ($C_{4v}$) symmetry. Eight elements of symmetry include, $m_z = I2_z$, $m_y$, $2_x$, $4_x$ and $m_{yz}$, where, for example, I denotes spatial inversion, $2_z$ is a dyad rotation about the z-axis (crystal c-axis) and $m_z$ is a mirror. All multipoles are charge-like. Axial ones have even rank, whereas there is no such restriction on polar multipoles. With signal enhancement provided by an electric dipole-electric dipole event (E1-E1) K = 0 and 2, while for a parity-odd electric dipole-electric quadrupole event (E1-E2) multipoles have ranks K = 1, 2, 3.

### III. RESONANT SCATTERING

The nature of the electronic ground state accessed in resonant Bragg diffraction depends on both the quantum labels of the virtual, intermediate core state and the type of the absorption event. Absorption at the ligand K-edge of transition metal compounds is directly related to covalency derived from ligand 2p-metal 3d hybridization [2]. In this picture, the wavefunction of the lowest unoccupied molecular orbital of a compound is represented by a linear combination of metal ($M_d$) and ligand ($L_p$) valence orbitals $[(1 - \rho^2)^{1/2} |M_d\rangle + \rho|L_p\rangle]$. The ligand K edge originates from a ligand 1s orbital and only intensities to the ligand component of the ground state wave function have intensity. The integrated intensity of the K pre-edge can be related to the covalence $\rho^2$ of the partially occupied 3d orbitals. Similarly, the metal $L_{2,3}$ transition 2p → 3d probes the empty metal 3d states, with an integrated intensity related to the covalence $(1 - \rho^2)$ of the partially unoccupied 3d orbitals. For instance, the 3d-transition metal oxides show large pre-edge structures in the oxygen K edge at an energy ≈ 0.53 keV; see, for example, experimental data for 11 compounds in Fig. 1 of Ref. [13]. E2 enabled transitions to the pre-edge can be excluded as the O 3d can be regarded as a continuum state. None of the oxides exhibit a strong pre-edge feature observed in many ligand Cl K-edge spectra [1, 2, 3, 4]. In the Ti K pre-edge of $TiO_2$ both E1 and E2 transitions are of importance and their separate contributions have been calculated by Uozumi *et al.*, cf. Fig. 3 in Ref. [14], and by Cabaret *et al.* [15]. While the dominant intensity arises from E1 transitions the evidence is that E2 transitions complete a satisfactory interpretation of the reported experimental data.

Assuming that virtual intermediate states are spherically symmetric, to a good approximation, the x-ray scattering length ≈ $\{F_{\mu\eta}/(E - \Delta + i\Gamma/2)\}$ in the region of the resonance, where $\Gamma$ is the total width of the resonance at an energy $\Delta$ [9, 16]. The numerator $F_{\mu\eta}$ is an amplitude, or unit-cell structure factor, for Bragg diffraction in the scattering channel with primary (secondary) polarization $\eta$ ($\mu$). By convention, $\sigma$ denotes polarization normal to the plane of scattering, and $\pi$ denotes polarization within the plane of scattering. Fig. 2 depicts polarization states, wave vectors, and the Bragg condition. The crystal is rotated through an azimuthal angle $\psi$ about the reflection vector $\kappa$.

Photon and electronic quantities in the scattering amplitude are partitioned in a generalized scalar product $F_{\mu\eta} = \{\mathbf{X}^K \cdot \langle \mathbf{O}^K \rangle\}$, with implied sums on rank K and associated projections Q in the interval $-K \leq Q \leq K$ [9, 17]. The complex conjugate of a multipole $\langle O^K_Q \rangle^* = (-1)^Q \langle O^K_{-Q} \rangle$, with a phase convention $\langle O^K_Q \rangle = [\langle O^K_Q \rangle' + i\langle O^K_Q \rangle'']$ for real and imaginary parts labelled by single and double primes, respectively. Selection rules on K and Q for the multipole imposed by symmetry of the site used by Cl are evidently duplicated in the x-ray factor $\mathbf{X}^K$. The latter is specific to a resonant event. One finds, $\mathbf{X}^K$ is independent of photon wave vectors for an E1-E1 event (K = 0 - 2) but this is not so for E1-E2 (K = 1 - 3). All information on x-ray factors needed here is found in Refs. [9, 17].

Returning to selection rules on Cl multipoles imposed by site symmetry, the element $m_z$ requires Q odd (even) for polar (axial), while a dyad element $2_x$ demands the relation $\langle O^K_Q \rangle = (-1)^K \langle O^K_{-Q} \rangle = (-1)^{K+Q} \langle O^K_Q \rangle^*$ is obeyed by both multipoles types. The requirement that projections are even or odd for axial and polar multipoles is a key factor in marked differences between corresponding scattering amplitudes, as we shall verify. A tetrad axis of rotation symmetry parallel to the x-axis is satisfied by a suitable linear combination of multipoles. Modelling in terms of standard spherical harmonics $C^K_Q$ reveals that real parts of $[C^2_0 - \sqrt{(6)} C^2_{+2}]$, $C^1_{+1}$ and $[C^3_{+1} - \sqrt{(5/3)} C^2_{+3}]$ are unchanged by $4_x$.

An electronic structure factor for diffraction embodies selection rules imposed by all elements of symmetry in a crystal space group. We use $\Psi^K_Q = [\exp(i\boldsymbol{\kappa} \cdot \mathbf{d}) \langle O^K_Q \rangle_\mathbf{d}]$ for a structure factor, where $\boldsymbol{\kappa}$ is defined by integer Miller indices $(h, k, l)$, and the implied sum is over all Cl sites $\mathbf{d}$ in a unit cell. Sites 24(e) in space group No. 225 are related by pure rotations alone, namely, $2_z$, $4_z$, $4_y$ [18] with the result,

$$\Psi^K_Q = [1 + (-1)^{h+l} + (-1)^{k+l} + (-1)^{h+k}] [\langle O^K_Q \rangle \{\alpha + (-1)^Q \alpha^* + \exp(i\pi Q/2)(\beta + (-1)^Q \beta^*)\}$$

$$+ (4^{-1}_y \langle O^K_Q \rangle) \gamma + (4^{+1}_y \langle O^K_Q \rangle) \gamma^*]. \qquad (1)$$

Extinction rules derived from the first factor in Eq. (1) include $h + l, k + l, h + k, = 2n$. There are no space-group forbidden reflections and allied Templeton-Templeton scattering. Spatial phase factors $\alpha = \exp(2\pi i x h)$, $\beta = \exp(2\pi i x k)$ and $\gamma = \exp(2\pi i x l)$ with integer Miller indices $h$, $k$, $l$, and general coordinate $x \approx 0.238$ [11]. Notably, axial and polar structure factors are functions of $\cos(2\pi x k)$ and $\sin(2\pi x k)$, respectively, and likewise for $\beta$ and $\gamma$. The identity $4^{-1}_y m_y \langle O^K_Q \rangle = 4^{-1}_y \langle O^K_Q \rangle = I4^{+1}_y \langle O^K_Q \rangle$ is useful, where the first equality follows from the invariance of Cl multipoles with respect to the mirror operation $m_y$ normal to the crystal b-axis.

### A. E1-E1

Scattering by spherical distributions of electronic charge observed in Thomson scattering is forbidden in the rotated channel of polarization ($\pi'\sigma$) [9, 17]. In practice, measured ($\pi'\sigma$) intensity might be contaminated by leakage of this strong contribution to unrotated channels of polarization ($\sigma'\sigma$) and ($\pi'\pi$).

Our calculations show that $(\pi'\sigma)_+ = 0$ for reflections of the type $(h, 0, 0)$ and $(0, 0, l)$, and also $(l, l, l)$. A subscript $+$ ($-$) attached to a scattering amplitude denotes E1-E1 (E1-E2). The dyad axis of rotation symmetry parallel to face diagonals in Cl site symmetry creates a two-fold harmonic in the azimuthal angle, e. g.,

$$(\pi'\sigma)_+ = 4\sqrt{6} \sin(\theta) \sin(2\psi) [1 - \cos(2\pi x k)] \langle T^2_0 \rangle; \qquad (0, k, k) \qquad (2)$$

Here, $\sin(\theta) \approx 0.319\ k$. The crystal a-axis is parallel to $-z$ in Fig. 2 at the azimuth origin $\psi = 0$. The result Eq. (2) for $(\pi'\sigma)_+$ with $k$ replaced by $h$ holds at $(h, h, 0)$, while for $\psi = 0$ the c-axis is normal to the plane of scattering (z-axis in Fig. 2). In Eq. (2), $\{\sin(\theta) [1 - \cos(2\pi x k)]\} \approx 0.295, 1.269, 1.171$ for $k = 1, 2, 3$, respectively.

### B. E1-E2

All polar scattering amplitudes vanish for $h = k = l = 0$, because the crystal is centrosymmetric. However, Cl ions can possess a polar dipole that contributes to scattering, as we shall verify. The identity $4^{+1}{}_y \langle U^K_Q \rangle = 0$ holds because projections Q of a polar multipole are odd, in order to satisfy the mirror $m_z$, and it simplifies $\Psi^K_Q$ in Eq. (1). Site symmetry demands multipoles are purely real, as for axial multipoles. While $(\pi'\sigma)_- = 0$ for Bragg reflections indexed $(h, 0, 0)$ scattering amplitudes indexed $(h, h, 0)$ and $(0, k, k)$ can be different from zero and they are not the same, unlike axial E1-E1. We find,

$$(\pi'\sigma)_- = (16/\sqrt{3}) \sin(2\pi x h) \sin^2(\theta) \sin(2\psi) \langle U^3_{+1} \rangle, \qquad (h, h, 0) \qquad (3)$$

$$(\pi'\sigma)_- = -(4/\sqrt{3}) \sin(2\pi x k) [(3/5) \sin(2\theta) \sin(\psi) \{\langle U^1_{+1} \rangle + 2\langle U^3_{+1} \rangle\}$$

$$+ 2\sin^2(\theta) \sin(2\psi) \langle U^3_{+1} \rangle]. \qquad (0, k, k) \qquad (4)$$

While the rotated amplitude for $(h, h, 0)$ in Eq. (3) displays simple two-fold periodicity in the azimuthal angle found for axial E1-E1 the same periodicity is possibly absent at reflections indexed by $(0, k, k)$. For the latter, the rotated amplitude Eq. (4) is proportional to $\sin(\psi)$ and zero at the origin where the crystal a-axis is parallel to $-z$ in Fig. 2. Intensity $|(\pi'\sigma)_-|^2$ derived from Eq. (4) as a function of $\psi$ is displayed in Fig. 3. For this exercise, Miller index $k = 1$ and 3, and the dipole $\langle U^1_{+1} \rangle = 0$. The corresponding intensity for $k = 2$ is made extremely small by the spatial phase factor $\sin(2\pi x k)$.

The unrotated amplitude $(\sigma'\sigma)_-$ for $(h, 0, 0)$ is predicted to be independent of the azimuthal angle. This is not so with $(h, h, 0)$ for which we find,

$$(\sigma'\sigma)_- = (16/(5\sqrt{3})) \sin(2\pi x h) \sin(\theta) [-3\langle U^1_{+1} \rangle$$

$$+ \{1 - 5 \cos(2\psi)\} \langle U^3_{+1} \rangle]. \quad (h, h, 0) \quad (5)$$

Amplitudes in Eqs. (4) and (5) include the Cl polar dipole $\langle U^1_x \rangle = -\sqrt{2} \langle U^1_{+1} \rangle$ directed along the crystal a-axis.

### IV. DISCUSSION

Motivated by a bold imprint from covalent bonding in the x-ray absorption spectrum in the vicinity of the Cl K-edge we report corresponding Bragg diffraction patterns for $K_2RuCl_6$ [1-5]. Our calculations include axial (E1-E1) and polar (E1-E2) absorption events, and they reveal the availability from resonant diffraction of significant information on bonding in the unusual compound [5]. Electronic properties are couched in terms of Cl multipoles that possess some Ru character through hybridization of Cl-p and Ru-d orbitals. Diffraction patterns are found to include contributions from a Cl polar dipole, although the crystal structure is centrosymmetric and not polar. The intense low energy peak of interest in the Cl absorption spectrum is at ≈ 2.820 keV. The Ru $L_3$-edge at ≈ 2.844 keV might restrict the quality of diffraction data.

The mineral Javorieite ($KFeCl_3$) adopts a structure quite different from cubic $K_2RuCl_6$, and it is meaningful to compare and contrast diffraction properties. For one thing, K and Fe resonance energies are well-separated from the Cl K-edge. Ferrous ($Fe^{2+}$, $3d^6$) ions in $KFeCl_3$ are bonded to six $Cl^{1-}$ ions to form edge-sharing octahedra in orthorhombic Pnma (No. 62), which is centrosymmetric like $Fm\bar{3}m$ used by $K_2RuCl_6$. All ions in Javorieite occupy special positions 4(c) with mirror symmetry $m_y$ ($C_s$) parallel to the crystal b-axis. Cell lengths are $a ≈$ 8.715 Å, $b ≈ 3.845$ Å, $c ≈ 14.15$ Å [19]. The Fe-Cl bond distance is about 1 Å shorter than the K-Cl distance. Isostructural copper halides have been gaining increased attention as highly luminescent nontoxic alternatives to lead halide perovskites for optoelectronic applications. Compounds include $K_2CuCl_3$ [20] and $Rb_2CuCl_3$ [21] with all elements in sites 4(c) in Pnma, just like Javorieite. $[CuCl_3]^{2-}$ chains are separated by $K^+$ cations in the former copper halide, for example.

Unlike cubic $K_2RuCl_6$ there are space-group forbidden reflections of the type ($h$, 0, 0) and (0, 0, $l$) with $h$, $l$ odd integers in diffraction by orthorhombic Javorieite at which Thomson scattering from spherical charge distributions is absent. Such scattering may contribute to the unrotated scattering amplitudes $(\sigma'\sigma)_+$ and $(\pi'\pi)_+$, but it is absent for $h$, $l$ odd that we consider. Templeton-Templeton scattering [22] using an E1-E1 event reveals a Cl quadrupole of xz-like spatial symmetry. Specifically, for ($h$, 0, 0) and an axial E1-E1 absorption event $(\sigma'\sigma)_+ = (\pi'\pi)_+ = 0$ and with $h = 1$ or 3,

$$(\pi'\sigma)_+ = -4\cos(\theta)\cos(\psi)\cos(2\pi xh)\langle T^2_{+1}\rangle'. \quad (h, 0, 0) \quad (6)$$

A direct comparison with the null result for $K_2RuCl_6$ must allow for absence of space-group forbidden reflections in the latter cubic compound. Amplitudes for rotated polarization and polar E1-E2 event include a Cl dipole parallel to the c-axis, while unrotated amplitudes are zero $(\sigma'\sigma)_- = (\pi'\pi)_- = 0$. For ($h$, 0, 0) with $h$ an odd integer $(\pi'\sigma)_- \propto \sin(2\pi xh)\sin(2\theta)\cos(\psi)$. Notably, the dependence of amplitudes on the azimuthal angle $\psi$ is the same for axial and polar absorption events. There is a different general coordinate x for each of the three independent Cl sites, namely, x ≈ 0.273, 0.171, 0.024 [19]. Chlorine multipoles in $(\pi'\sigma)_-$ include the dipole $\langle U^1_0\rangle$ and a linear combination of $\langle U^2_{+2}\rangle''$, $\langle U^3_0\rangle$, $\langle U^3_{+2}\rangle'$. Similar results are derived for (0, 0, $l$) and $l$ = 1, 3 or 5 with general coordinates z ≈ 0.202, 0.491, 0.903 [19]. All results mentioned for Javorieite apply to $K_2CuCl_3$ and $Rb_2CuCl_3$ on using appropriate cell lengths and general coordinates [20, 21].

Lastly, we mention a cupric halide with a chiral structure. The compound $CsCuCl_3$ adopts the enantiomorphic space-group pair $P6_122$ (No. 178) or $P6_522$ (No. 179) at room temperature, with copper ions disturbed from perfect octahedral coordination by a tetragonal distortion [23]. Chlorine ions use sites 12(c) with no symmetry and special positions 6(b) that contain a dyad axis of rotation symmetry. Structural chirality has been confirmed by resonant diffraction of helical x-rays [24], together with the observation of magneto-chiral dichroism in a strong magnetic field (14.5 T) at T = 4.2 K [25].

**Acknowledgement** Dr D. D. Khalyavin prepared Fig. 1. Professor S. P. Collins advised on the feasibility of the proposed resonant x-ray Bragg diffraction.

-------------------------------------------------------------------------------------------------

**TABLE I**. Atomic radial integrals for E1 and E2 absorption events of the Cl, Ti, and Ni K-edges in units of the Bohr radius [8]. The magnitude of E1 (E2) decreases (increases) with resonance energy $\Delta$.

Cl, $\Delta \approx 2.82$ keV
E1; $(1s\|R\|3p) = -0.01776$: E2; $(1s\|R^2\|3d) = 0.00244$
E2/E1 = $-0.137$

Ti, $\Delta \approx 4.97$ keV
E1; $(1s\|R\|4p) = -0.00618$: E2; $(1s\|R^2\|3d) = 0.00129$
E2/E1 = $-0.209$

Ni, $\Delta \approx 8.34$ keV
E1; $(1s\|R\|4p) = -0.00294$: E2; $(1s\|R^2\|3d) = 0.00082$
E2/E1 = $-0.279$

-------------------------------------------------------------------------------------------------

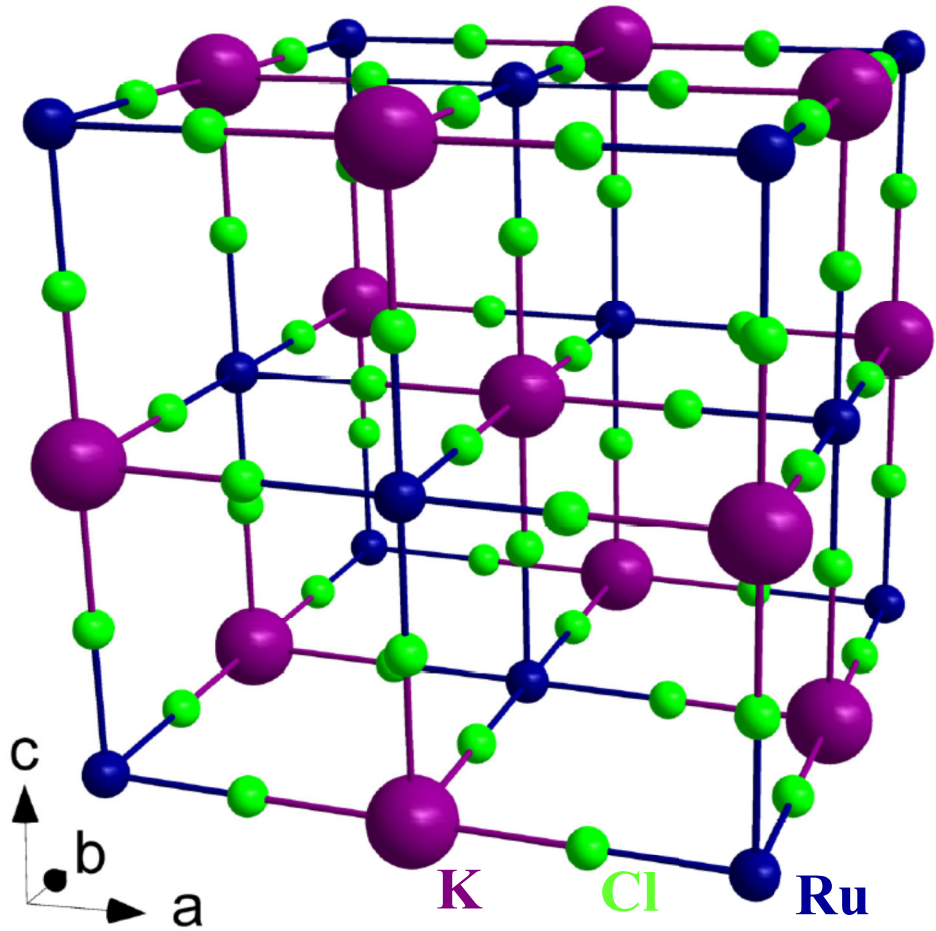

FIG. 1. Crystal structure of K$_2$RuCl$_6$ [11]. Cell length $a \approx 9.737$ Å.

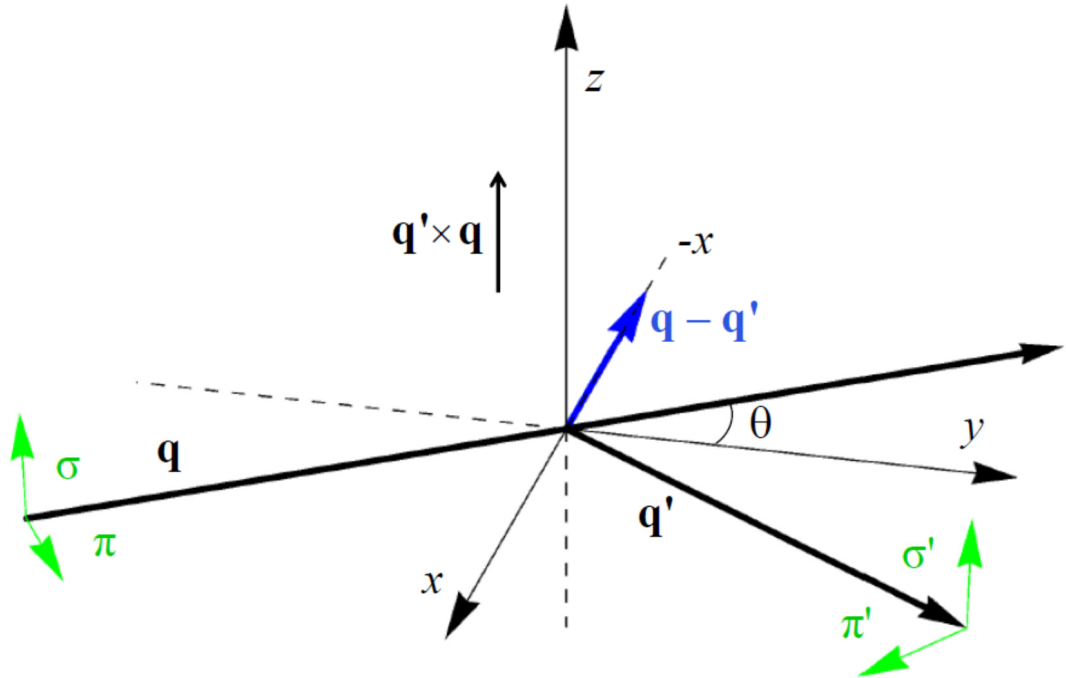

FIG. 2. Primary ($\sigma$, $\pi$) and secondary ($\sigma'$, $\pi'$) states of polarization. Corresponding wave vectors **q** and **q**′ subtend an angle $2\theta$, and the reflection vector $\boldsymbol{\kappa} = \mathbf{q} - \mathbf{q}'$.

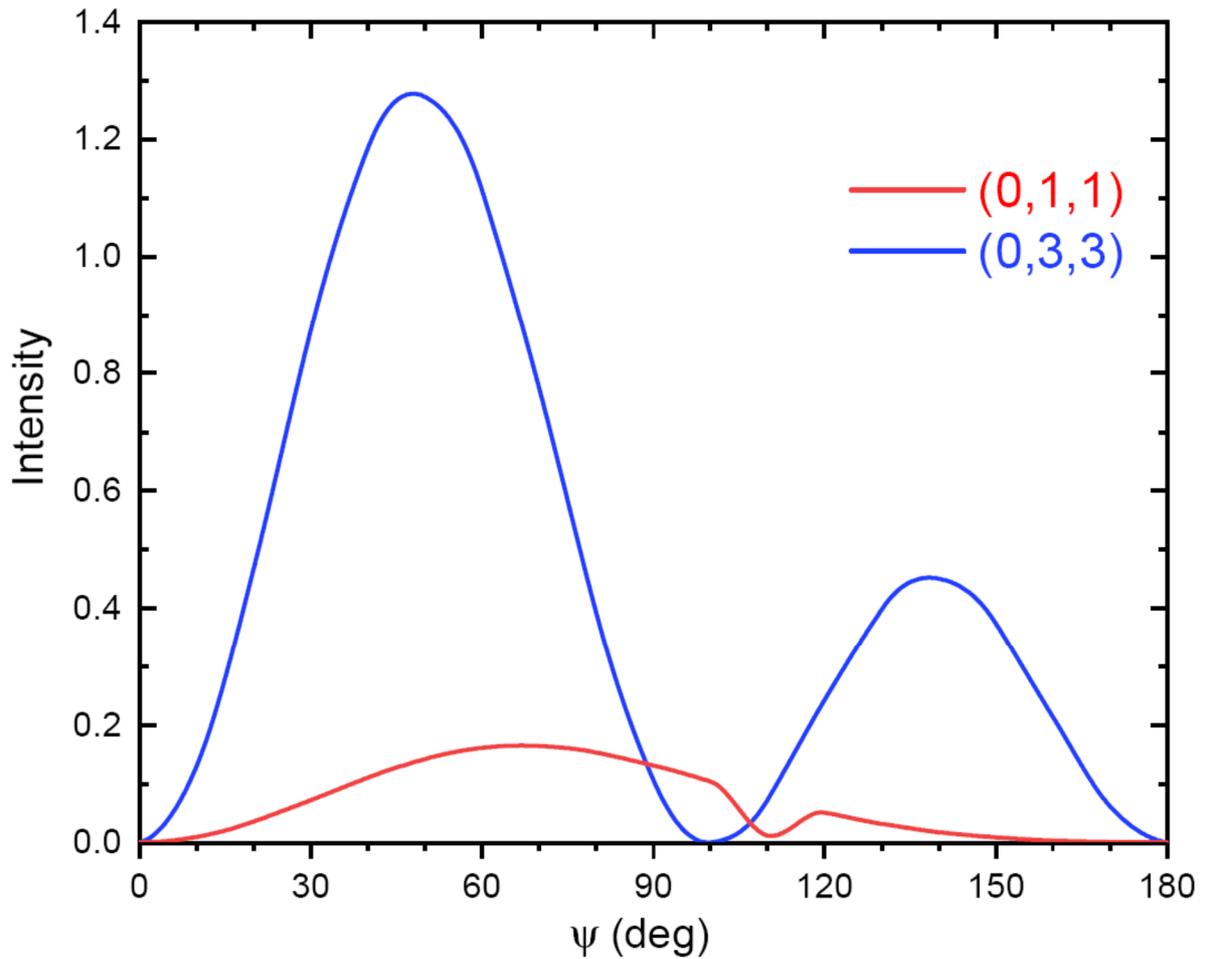

FIG. 3. Intensity $|(\pi'\sigma)_-|^2$ of Bragg spots indexed $(0, k, k)$ as a function of $\psi$ derived from the amplitude in Eq. (4) for $K_2RuCl_6$. Red $(0, 1, 1)$; Blue $(0, 3, 3)$, with the polar dipole $\langle U^1_x \rangle = -\sqrt{2} \langle U^1_{+1} \rangle$ arbitrarily set to zero. In consequence, the scale of $|(\pi'\sigma)_-|^2$ is set by $\langle U^3_{+1} \rangle^2$, and $\sin^2(2\pi x k)$ is included in the displayed quantities.